# Ferroelectricity
# of Néel-type magnetic domain walls

A.P. Pyatakov[*], D.A. Sechin, A.V. Nikolaev, E.P. Nikolaeva, A.S. Logginov

Physics Department, M.V. Lomonosov Moscow State University, Leninskie gory, Moscow, 119991, Russia

*) E-mail: pyatakov@physics.msu.ru

The chirality-dependent magnetoelectric properties of Néel-type domain walls in iron garnet films is observed. The electrically driven magnetic domain wall motion changes the direction to the opposite with the reversal of electric polarity of the probe and with the chirality switching of the domain wall from clockwise to counterclockwise. This proves that the origin of the electric field induced micromagnetic structure transformation is inhomogeneous magnetoelectric interaction.

## 1. Introduction

In recent years there has been the remarkable progress in understanding of the coupling mechanisms between magnetic and electric subsystems in multiferroics [1-11]. A lot of excitement was caused by the discovery of so called *spiral multiferroics* in which ferroelectricity is induced by spatially modulated spin structure since magnetoelectric coupling expected to be particularly strong in these media [12-17]. The chirality of the spin spiral determines the direction of polarization, as was proved experimentally by reversal the chirality with electric field in orthorhombic manganites $RMnO_3$ (R=Dy, Tb) [18; 19] and $MnWO_4$ [20]. The specific type of ferroelectric domains generated by chiral structures was observed in [21,22]. This magnetically induced ferroelectricity is caused by the special type of magnetoelectric interaction, the *inhomogeneous* one, that is described by $P_i M_j \nabla_k M_n$ coupling term, where P and M, are electric and magnetic order parameters, respectively [23-26].

Micromagnetic structures like domain walls and magnetic vortexes can be also considered as a source of ferroelectricity due to the modulation of magnetic order parameter that occurs in them. Noteworthy that the term "inhomogeneous magnetoelectric interaction" was originally used by V.G. Bariakhtar [25] in relation to the domain wall in magnet. In subsequent works [27] the magnetoelectric properties of antiferromagnetic domain were discussed. This topic recently was revitalized by I.E. Dzyaloshinskii [28] that proposed the nucleation of domain walls with electric field and its motion in the gradient of electric field. It should be noted that ferroelectricity developed from micromagnetic structure is universal phenomena and can appear in every magnetic insulator even in centrocymmetric one [28]. Quite surprisingly the experimental proof for electric properties of magnetic domain walls is still lacking. The indirect evidence can be seen in the enhancement of electromagnetooptical effect (i.e. electric field induced Faraday rotation [29]) in the vicinity of domain wall observed in YIG films [30,31].

Recently the motion of domain walls in the gradient electric field provided by a tip electrode was observed in rare earth iron garnet films [32-34]. Although the basic features of the effect agree with the properties of inhomogeneous magnetoelectric interaction $P_i M_j \nabla_k M_n$ (i.e. the oddness with respect to the electric polarity and evenness with respect to the magnetic polarity of the domain), another mechanism of the effect is possible such as the electrically induced local variation of magnetic anisotropy [30,31] that forces the domain walls to be attracted or repelled by the tip. Thus the origin of the effect have remained unclear since the key feature of inhomogeneous magnetoelectric interaction, namely, the dependence on the domain wall chirality has not been shown.

In this paper the magnetoelectric properties of Néel-type domain walls with controlled chirality are studied in iron garnet film. The electrically induced domain wall displacement that depends on the chirality of the walls shows unambiguously that the origin of the effect is inhomogeneous magnetoelectric interaction.

## 2. Theory

The inhomogeneous magnetoelectric effect corresponds to the following contribution to the free energy of the crystal:

$$F_{ME} = \gamma_{ijkl} \cdot P_i \cdot M_j \cdot \nabla_k M_l, \qquad (1)$$

where **M**=**M(r)** is magnetization distribution, **P** is electric polarization, $\nabla$ is vector differential operator, $\gamma_{ijkl}$ is the tensor of inhomogeneous magnetoelectric interaction that is determined by the symmetry of the crystal.

The inhomogeneous magnetoelectric contributions (1) for the bulk crystal of iron garnets with cubic symmetry takes the following high symmetry form [26,13]:

$$F_{ME} = \gamma \cdot \mathbf{P} \cdot (\mathbf{M} \cdot (\nabla \cdot \mathbf{M}) - (\mathbf{M} \cdot \nabla)\mathbf{M}). \qquad (2)$$

Taking into account that operator $\nabla$ can be represented in terms of magnetic spiral wave vector **k** as $\mathbf{k}\frac{\partial}{\partial x}$, where $x$ is the coordinate along the axis parallel to **k** the following form for (2) can be obtained:

$$F_{ME} = \gamma\left(\mathbf{P}\cdot\left[\mathbf{k}\times\left[\frac{\partial \mathbf{M}}{\partial x}\times\mathbf{M}\right]\right]\right) = \gamma M^2(\mathbf{P}\cdot[\mathbf{k}\times\mathbf{\Omega}]) \qquad (3)$$

where **Ω** is the spin rotation axis. Thus in accordance with the rule formulated in [13,6], the electric polarization of the spatially modulated structure can be found as a vector product of **k** and **Ω**:

$$\mathbf{P} = -\frac{\partial F_{ME}}{\partial \mathbf{E}} = \gamma \chi_e M^2 [\mathbf{\Omega}\times\mathbf{k}] \qquad (4)$$

where $\chi_e$ is electric susceptibility.

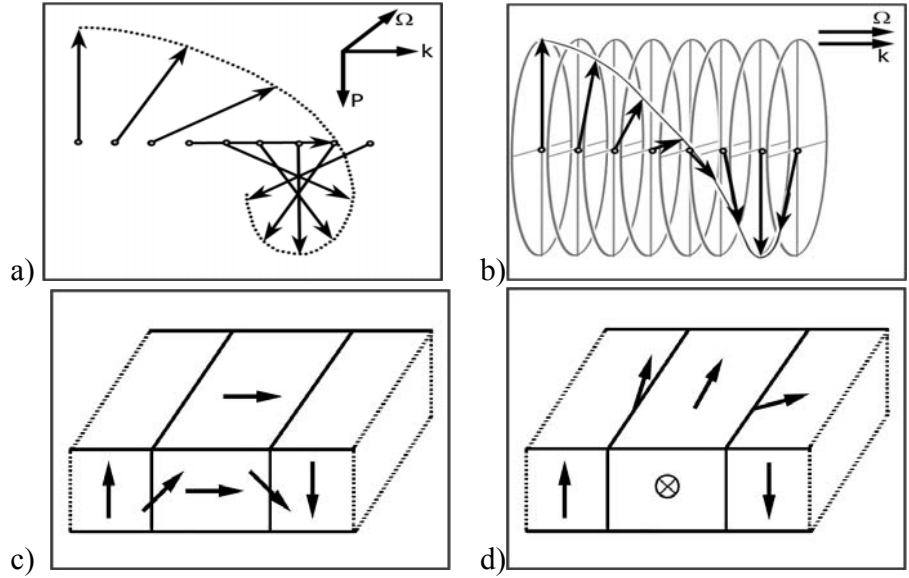

Fig. 1. Spatially modulated magnetic structures in media a) a spin cycloid structure b) a spin helicoid structure c) a Néel domain wall d) a Bloch domain wall.

It follows from (4) that the magnetically induced polarization is nonvanishing for cycloidal modulation (fig.1 a), i.e. the spiral plane is parallel to the modulation wave vector and should be zero for helicoidal modulation in which **k**||**Ω** (fig. 1 b). One can see that domain wall of Néel type can be considered as a part of the spin cycloid (fig.1 c) and the Bloch-type domain wall as a part of helicoid (fig. 1d). It follows also from (4) that upon changing the chirality of the cycloid or Néel domain wall (**Ω** ⇨ -**Ω**) the electric polarization of the spatially modulated structure reverses (**P** ⇨ - **P**).

In the bulk of the magnetic media the domain walls of Bloch type correspond to the state with the lowest energy since magnetic charges at the boundary are zero (div**M**=0). However in the magnetic field perpendicular to the domain wall plane the Bloch domain wall transforms into Néel type one [35]. In the following experimental part we report the magnetoelectric properties of domain wall in the Néel-type state with definite chirality induced by magnetic field. The effect of electric polarization reversal upon chirality switching is demonstrated.

## 3. Experiment

Iron garnet films are well known magnetooptical materials, thus one can visualize domains and domain walls, for example with Faraday effect [36, 37]. The experimental scheme is shown in Figure 2. To induce the Néel-type state of the domain wall the in-plane magnetic field produced by magnetic coils was applied perpendicular to the domain wall. To probe the electrical properties of the domain wall the tip electrode was used. It was made of copper wire that was sharpened by etching in ferric chloride to curvature radius of ~2 μm at one end. Such probe allowed us to obtain an electric field with strength of up to 7 MV/cm near the point of contact by supplying voltage up to 1500 V. Strong electrostatic field did not cause dielectric breakdown because it decreased rapidly with the distance from the probe and near the grounding electrode (i.e. metal foil attached to substrate) the strength of the field did not exceed 100 V/cm.

In our experiment we used $(BiLu)_3(FeGa)_5(O)_{12}$ films grown by liquid-phase epitaxy on (210) $Gd_3Ga_5O_{12}$ substrate (for details see [38]). The parameters of the samples are listed in Table 1. Samples with (210) substrate orientation were chosen because the anisotropy in them is strong enough to retain the orientation of the domain wall along $[\bar{1}20]$ axis even in the presence of magnetic field normal to the domain wall plane, while in highly symmetrical (111) films the domain walls align along the field thus making the induction of Néel-state impossible.

| Sample no. | $h(\mu m)$ | $4\pi M_s(Gs)$ | $p(\mu m)$ |
|---|---|---|---|
| 1 | 9.9 | 62.37 | 28.11 |
| 2 | 18.7 | 62.17 | 26.25 |
| 3 | 11 | 43.75 | 35.6 |

Table 1: Parameters of the samples under study. Symbol $h$ stands for sample thickness, $M_S$ is the saturation magnetization, and $p$ is a period of domain structure [38].

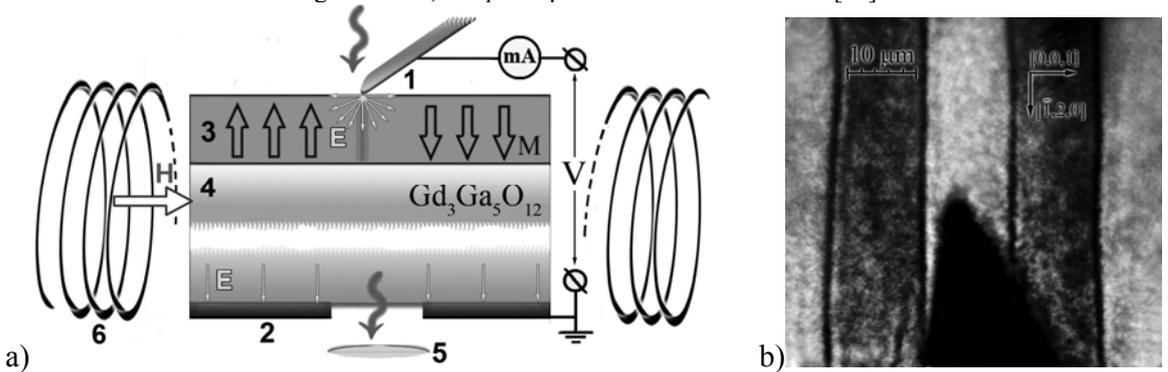

a)          b)

Fig. 2. a) Schematic representation of the geometry of the experiment and the configurations of the magnetic field, electric field and magnetization. The electric field is formed in the dielectric medium of the sample between the tip (1) and the copper foil (2), which plays the role of the grounding electrode, (3) is iron garnet film, (4) is the GGG substrate, (5) is optical system, (6) are magnetic coils.
b) The top view magnetooptical image of the domain structure in iron garnet film: dark and bright gray stripes are domain with upward and downward direction of magnetization, respectively. Vertical black lines are domain wall images, dark area at the bottom is the image of electrode tip.

## 4. Results

In experiment we registered magnetic structure in the initial state (fig. 2 b) and after applying external electric and magnetic fields (fig. 3).

The application of in-plane magnetic field of several tens Oe without electric field did not lead to significant changes of the micromagnetic structure while application electric voltage of 500÷1500V between the tip and the ground electrode caused the evident displacement of the domain wall. Its direction depends on both the polarity of the field and the magnetic field direction (fig. 3). As one can see on fig. 3 the change in magnetic field direction causes reversal of the sign of wall displacement. It is important to note simultaneous movement of the wall that is nearest to the probe and of the next wall to the right: while one is attracted the other is repulsed and vice versa.

It worth mentioning that the effect of the domain wall displacement was observed in the samples listed in Table 1 even in the absence of magnetic field (for details, see [33]), however its value was much smaller and the positive voltage always caused the attraction of the domain wall nearest to the tip, while negative voltage always caused the repulsion irrespective of the position of the tip.

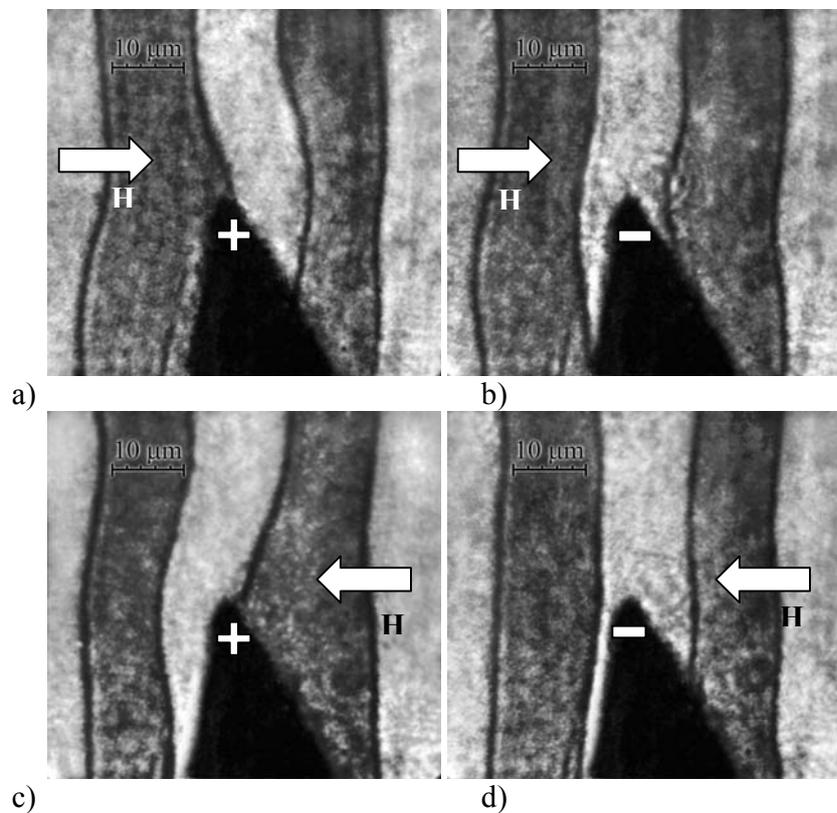

Fig. 3. The transformation of the micromagnetic structure in the magnetic and electric fields applied simultaneously: a) the positive potential of the tip, and LTR direction of magnetic field b) negative potential at the tip, LTR direction of magnetic field; c) the positive potential of the tip, and RTL direction of magnetic field d) the negative potential of the tip, and RTL magnetic field. The sample 1 from the table is used.

Summarizing the experimental results we can point out the following characteristic features of the electric field driven domain wall motion:

(i) Switching the electric polarity changes the sign of the effect
(ii) Neighboring walls shows opposite direction of the displacement for the same polarity of electric field
(iii) Switching the magnetic polarity of the in-plane field changes the sign of the effect

## 5. Discussion

The electric field driven domain wall motion can be explained by ferroelectricity of the walls in accordance with the predictions [25,14,28]. However one can not rule out another scenario according to which the electric field induces variation of magnetic anisotropy [30,31]. In the immediate vicinity of the contact point the local anisotropy can favor (unfavor) the in-plane direction of the magnetization, resulting, respectively, in the attraction (repulsion) of the domain wall. Thus the discriminative feature of spin spiral induced ferroelectricity is the polarization reversal upon the switching chirality of the spiral. In the following we will show that properties of the effect (i-iii) are explained in terms of inhomogenous magnetoelectric interaction (2) and magnetically induced electric polarization (4).

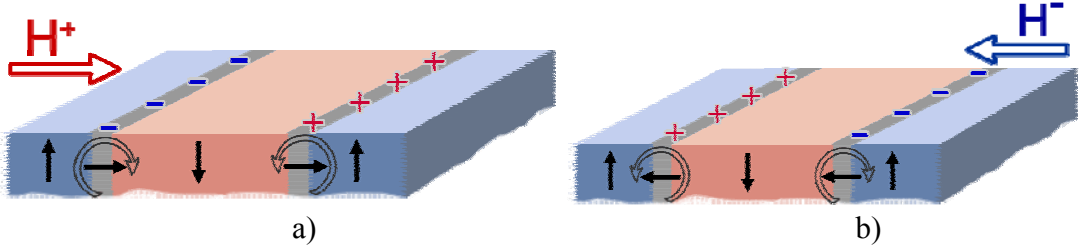

a)          b)

Fig. 4 The schematic representation of micromagnetic structures corresponding to the experimental images of Fig. 3: (a) corresponds to the a, b of Fig.3 (b) corresponds to the c, d of Fig.3. For illustrative purposes the size of domain wall is exaggerated.

Let us consider the stripe domain structure in magnetic film (fig.4). If we choose the modulation vector **k** directed left to right than the left boundary of the central domain is characterized by the clockwise direction of magnetization rotation while neighboring domain boundary has the opposite chirality (fig 4 a). The corresponding electrical charges due to inhomogeneous magnetoelectric interaction (formula 4) are shown at the domain wall.

When the in-plane magnetic field is reversed the chirality and hence the electric polarity of the charges change the sign (fig 4 b) in accordance with the feature (iii).

It is easily seen from the picture (4 b) that neighboring walls have opposite chiralities: magnetization rotates clockwise and counterclockwise. The chirality determines the direction of electric polarization in the wall so the neighboring walls should posses opposite electric polarities in accordance with feature (ii). The later makes us rule out the effects related to the local anisotropy changes as in the single crystal the anisotropy is the same for all points of the sample and should not depend on the domain wall chirality.

The results agree with the equation (4) and the results of the theoretical work [27] in which the antiferromagnetic domain wall in electric and magnetic field applied simultaneously are considered.

Finally some comments should be made about the effect of domain wall displacement that was observed in the same samples even in the absence of magnetic field [32-34]. As this magnetic field induces Néel-type state of domain wall the question arises how the electric field alone can induce attraction or repulsion of the domain wall. We relate it with some "spontaneous" Néel component of domain wall that arises due to the fact that direction of magnetization in the domain is not exactly parallel to the domain wall plane. It has the projection on the direction of modulation vector thus making nonzero terms in (2) $\mathbf{M} \cdot (\nabla \cdot \mathbf{M}) \neq 0; (\mathbf{M} \cdot \nabla)\mathbf{M} \neq 0$ (for details see [33]).

Authors are grateful to A.K. Zvezdin for the interest to the work and valuable discussions.